\documentclass[sigconf]{acmart}

\usepackage{enumitem}

\usepackage{amssymb}
\usepackage{pifont}
\usepackage{mathtools}
\usepackage{url}
\usepackage{multirow}
\usepackage{subfigure}
\usepackage{bm}
\usepackage{verbatim}
\usepackage{ragged2e}
\usepackage{caption}
\usepackage{subcaption}
\usepackage{graphicx}
\usepackage[normalem]{ulem}
\useunder{\uline}{\ul}{}
\usepackage{amsmath}
\usepackage{longtable}
\usepackage{adjustbox}
\usepackage{microtype}
\usepackage{setspace}
\usepackage{hyperref}
\usepackage{float}
\usepackage{balance}
\usepackage{algorithm}
\usepackage{algpseudocode}

\newcommand{\aka}

\AtBeginDocument{%
  }

\copyrightyear{2026}
\acmYear{2026}
\setcopyright{cc}
\setcctype{by}
\acmConference[CIKM '26] {Proceedings of the 35th ACM International Conference on Information and Knowledge Management}{November 7--11, 2026}{Rome, Italy.}
\acmBooktitle{Proceedings of the 35th ACM International Conference on Information and Knowledge Management (CIKM '26), November 7--11, 2026, Rome, Italy}
\acmISBN{979-8-4007-2539-5/2026/11}
\acmDOI{10.1145/3799682.3840083}

\begin{document}

\title{MERGE: Next-Generation Item Indexing Paradigm for Large-Scale Streaming Recommendation}


\author{Jing Yan}
\orcid{0009-0001-7014-4750}
\affiliation{
  \institution{Bytedance}
  \city{Singapore}
  \country{Singapore}
}
\email{yanjing.rec@bytedance.com}
\authornote{Corresponding author.}

\author{Yimeng Bai}
\orcid{0009-0008-8874-9409}
\affiliation{
  \institution{Bytedance}
  \city{Shanghai}
  \country{China}
}
\email{baiyimeng@bytedance.com}

\author{Zongyu Liu}
\orcid{0000-0003-4351-9518}
\affiliation{
  \institution{Bytedance}
  \city{Singapore}
  \country{Singapore}
}
\email{liuzongyu@bytedance.com}

\author{Yahui Liu}
\orcid{0000-0002-5923-1511}
\affiliation{
  \institution{Bytedance}
  \city{Shanghai}
  \country{China}
}
\email{liuyahui.lydia@bytedance.com}

\author{Junwei Wang}
\orcid{0009-0009-8512-3678}
\affiliation{
  \institution{Bytedance}
  \city{Shanghai}
  \country{China}
}
\email{wangjunwei.00@bytedance.com}

\author{Jingze Huang}
\orcid{0009-0005-3087-0781}
\affiliation{
  \institution{Bytedance}
  \city{Shanghai}
  \country{China}
}
\email{huangjingze@bytedance.com}

\author{Haoda Li}
\orcid{0000-0002-4570-4728}
\affiliation{
  \institution{Bytedance}
  \city{Singapore}
  \country{Singapore}
}
\email{lihaoda@bytedance.com}

\author{Sihao Ding}
\orcid{0000-0003-1796-8504}
\affiliation{
  \institution{Bytedance}
  \city{Shanghai}
  \country{China}
}
\email{dingsihao@bytedance.com}

\author{Shaohui Ruan}
\orcid{0009-0003-0150-0445}
\affiliation{
  \institution{Bytedance}
  \city{Shanghai}
  \country{China}
}
\email{ruanshaohui@bytedance.com}

\author{Yang Zhang}
\orcid{0000-0002-7863-5183}
\affiliation{
  \institution{National University of Singapore}
  \city{Singapore}
  \country{Singapore}
}
\email{zyang1580@gmail.com}




\renewcommand{\shortauthors}{Jing Yan et al.}

\begin{abstract}
Item indexing, which maps a large corpus of items into compact discrete representations, is critical for both discriminative and generative recommender systems, yet existing Vector Quantization (VQ)-based approaches struggle with the highly skewed and non-stationary item distributions common in industrial streaming recommenders, leading to poor assignment accuracy, imbalanced cluster occupancy, and insufficient cluster separation. To address these challenges, we propose MERGE, a next-generation item indexing paradigm that adaptively constructs clusters from scratch, dynamically monitors cluster occupancy, and forms hierarchical index structures via fine-to-coarse merging. Extensive experiments demonstrate that MERGE significantly improves assignment accuracy, cluster uniformity, and cluster separation compared with existing indexing methods, while online A/B tests show substantial gains in key business metrics, highlighting its potential as a foundational indexing approach for large-scale recommendation. Codes are available at \url{https://github.com/baiyimeng/MERGE}.

\end{abstract}





\maketitle

\section{Introduction}\label{sec:intro}

\textit{Item indexing}, also referred to as item tokenization, which aims to map a large corpus of items into a compact, discrete set of indexes\footnote{The terms ``indexes'', ``tokens'', ``codes'', ``codewords'', and ``clusters'' are often used interchangeably in different contexts.}, plays a critical role in modern recommender systems~\cite{SID_survey,GRID}. In traditional discriminative recommendation~\cite{RankMixer}, it enables the construction of efficient index structures that can manage massive candidate sets while satisfying the low-latency requirements of the retrieval stage, thereby providing a solid foundation for the subsequent ranking stage~\cite{TDM,DR}. In emerging generative recommendation~\cite{GenRec_survey,TIGER}, it supports the direct generation of items within a compressed token space, thus determining the upper bound of downstream recommender model performance~\cite{BLOGER,DiscRec}. Consequently, item indexing has attracted substantial attention from both the academic and industrial communities~\cite{StreamingVQ,LETTER,SEATER}.

Early works on item indexing explored handcrafted tree-based~\cite{TDM,JTM,OTM} and hierarchical path~\cite{DR} index structures, achieving a breakthrough beyond classical two-tower retrieval frameworks. Motivated by the remarkable success of generative AI~\cite{Qwen3,deepseekv3}, recent studies have increasingly adopted Vector Quantization (VQ)-based tokenization techniques~\cite{VQ-VAE} for recommendation~\cite{StreamingVQ,Trinity,TIGER,MMQv2,OneRecv2,OneSearch,COBRA,CAT-ID}, which learn one or more codebooks that serve as the basis for item indexing. For instance, StreamingVQ~\cite{StreamingVQ} directly employs VQ for single-layer clustering to enhance index immediacy, effectively replacing all major online retrievers. Trinity~\cite{Trinity} utilizes VQ for two-level clustering to construct statistical interest histograms of users, providing a unified solution for modeling multi-interest, long-tail, and long-term interests.

Despite their widespread success, we argue that VQ techniques are poorly suited for industrial streaming recommendation scenarios. When faced with highly skewed and rapidly evolving item distributions~\cite{DROS}, VQ methods encounter several challenges.  Specifically, focusing on a single-layer VQ, \textit{i.e.}, a single codebook comprising multiple codewords (clusters), we identify three key challenges: (1) \textit{Accuracy}: Item embeddings are poorly aligned with their assigned cluster embeddings, with an average cosine similarity of only 0.6, potentially leading to reduced retrieval precision. (2) \textit{Uniformity}: Cluster occupancy is highly imbalanced, with some clusters containing orders of magnitude more items than others. This imbalance typically manifests in the indiscriminate grouping of long-tail items, thereby hindering the effective retrieval of high-quality long-tail candidates. (3) \textit{Separation}: Cluster embeddings are insufficiently distinct, exhibiting an average inter-cluster cosine similarity exceeding 0.5, which can result in training instability, \textit{i.e.}, the same item being assigned to different clusters over short periods. These challenges could be further exacerbated by feedback loops~\cite{FBLoop}, ultimately destabilizing the recommendation ecosystem.

This naturally raises a broader question: can we establish a new solution for item indexing, specifically designed for industrial recommendation, that overcomes these inherent limitations? Addressing these challenges calls for a rethinking of the fundamental principles of index construction. Regarding accuracy, when assigning an item to a cluster, the assignment should be rejected if the similarity is too low, and new clusters should be dynamically created when necessary. In terms of uniformity, adaptive mechanisms are needed to regulate cluster sizes within reasonable bounds, thereby mitigating severe occupancy imbalances. Regarding separation, cluster construction should be context-aware, taking the existence of other clusters into account to ensure adequate distinctiveness.

In response, we propose \textbf{MERGE}, a next-generation item indexing paradigm tailored for large-scale streaming recommendation. Unlike VQ, which depends on a predefined number of clusters, MERGE adaptively generates clusters entirely from scratch, thereby enabling the formation of a dynamic codebook that evolves with the streaming data. 
Within each updating step, 
items are initially matched to existing clusters based on similarity, with matches exceeding a threshold updating cluster embeddings via the Exponential Moving Average (EMA) algorithm~\cite{EMA}. Items that fail to match are processed using a Union-Find algorithm~\cite{UF} to form new clusters, while those insufficient to establish a cluster center are temporarily cached. Cluster occupancy is also dynamically monitored, retaining only clusters with an appropriate number of items and resetting those that do not meet this criterion. Overall, this approach---combining dynamic cluster construction with real-time occupancy monitoring---overcomes the aforementioned limitations. Building on this, we treat these clusters as a fine-grained codebook and adaptively aggregate it into coarser-grained codebooks, ultimately forming a hierarchical indexing structure. We conduct both offline and online evaluations, yielding solid evidence that attests to its superior recommendation performance.

The main contributions of this work are summarized as follows:
\begin{itemize}[leftmargin=*]
\item We identify three fundamental challenges in existing item indexing techniques for large-scale streaming recommendation---accuracy, uniformity, and separation---highlighting the limitations of VQ-based approaches in handling highly skewed and non-stationary item distributions. 
\item We propose MERGE, a next-generation item indexing paradigm that adaptively generates clusters from scratch, dynamically monitors cluster occupancy, and builds a hierarchical indexing structure via fine-to-coarse merging, providing a principled solution tailored for large-scale recommendation.
\item We conduct extensive offline evaluations and online deployments, demonstrating that MERGE consistently outperforms baseline indexing methods and yields significant improvements across multiple business metrics. 
\end{itemize}
\section{Preliminaries}

\subsection{Task Formulation}

Item indexing maps items into compact discrete representations, enabling efficient index structures for retrieval in multi-stage recommendation pipelines~\cite{TDM,DR}. These representations can also serve as token spaces for generative recommendation, where recommendation is formulated as sequence generation~\cite{OneRecv2}.

Each item $i$ is represented by an embedding $\bm{e}_i$, obtained from either a pretrained recommender model~\cite{ETEGRec} or a multimodal model such as Qwen~\cite{Qwen3}. Item indexing learns a mapping:
\begin{equation}\label{eq:indexing}
f :  \bm{e}_i \rightarrow [c^1, c^2,\dots,c^L].
\end{equation}
Here, $f$ assigns $\bm{e}_i$ a discrete index of $L$ codewords. In traditional retrieval, $L$ is usually 1 or 2, allowing multiple items to share an index~\cite{Trinity,StreamingVQ}. Generative recommenders often use longer indices (\textit{e.g.}, $L=3$ or $4$), with an additional unique code to ensure one-to-one item--index correspondence~\cite{ETEGRec,TIGER}.

\subsection{VQ-Based Indexing Paradigm}

We first consider single-layer VQ and omit VAE-based extensions for simplicity~\cite{VQ-VAE}. In this case, \(L=1\), and the discrete representation space is a learnable codebook \(\mathcal{C}= \{\bm{c}_k\}_{k=1}^{K}\), where \(K\) is the codebook size and each codeword \(\bm{c}_k\) has the same dimension as \(\bm{e}_i\). Each item is assigned to the nearest codeword under a similarity metric \(\text{sim}(\cdot, \cdot)\), typically the negative squared Euclidean distance:
\begin{equation}\label{eq:vq}
c^1 = \mathop{\arg\max}\limits_{{k \in \{1, \dots, K\}}} \text{sim}(\bm{e}_i, \bm{c}_k).
\end{equation}

This formulation extends naturally to multi-layer residual quantization (RQ)~\cite{TIGER}, which uses coarse-to-fine residual dependencies with codebooks \(\mathcal{C}^l = \{\bm{c}_k^l\}_{k=1}^{K}\), \(l=1,\dots,L\). The \(l\)-th quantization layer is defined as:
\begin{equation}\label{eq:rq}
\begin{aligned}
c^l &=  \mathop{\arg\max}_{k \in \{1, \dots, K\}} \text{sim}(\bm{r}^l, \bm{c}_k^l), \\
\bm{r}^l &= \bm{r}^{l-1} - \bm{c}_{c^l}^l, \\
\bm{r}^0 &= \bm{e}_i,
\end{aligned}
\end{equation}
where \(\bm{c}_{c^l}^l\) is the selected codeword from the \(l\)-th codebook, \(\bm{r}^l\) is the residual after layer \(l\), and \(\bm{r}^0\) is the original item embedding.

\section{Methodology}\label{sec:method}
\begin{figure*}[t]
    \centering
    \includegraphics[width=0.9\linewidth, height=0.46\linewidth]{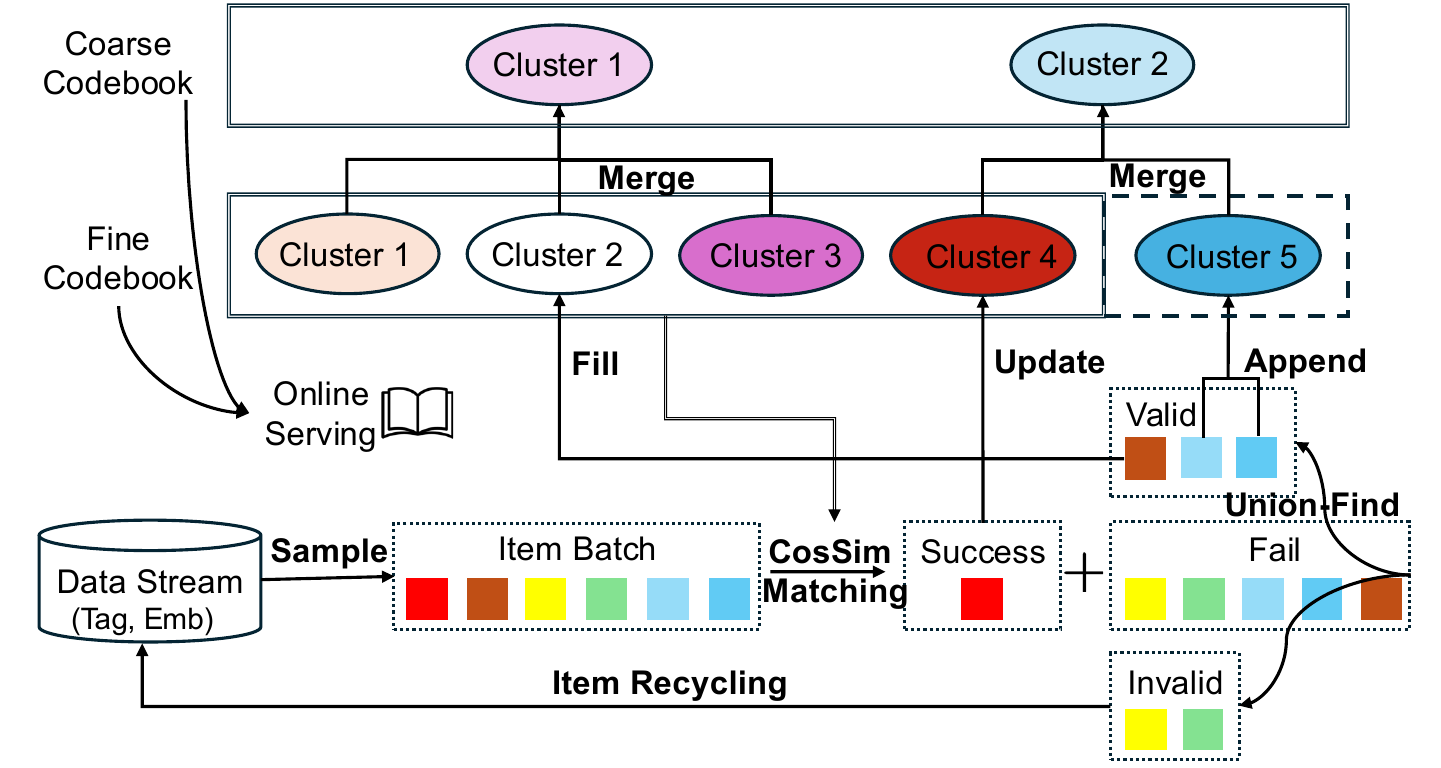}
    \caption{Overview of MERGE. Incoming item batches are matched to the codebook via cosine similarity. Matched items update existing clusters, while unmatched items form new clusters through union-find. Valid new clusters are filled or appended, and the rest are recycled. A fine-to-coarse merging process further builds a hierarchical codebook for efficient indexing.}
    \label{fig:framework}
\end{figure*}\textit{}
\subsection{Overview}

We propose \textbf{MERGE}, an item indexing method for industrial streaming recommendation, to address the three challenges in Section~\ref{sec:intro}: \textit{accuracy}, \textit{uniformity}, and \textit{separation}. As shown in Figure~\ref{fig:framework}, MERGE dynamically organizes incoming items into clusters while keeping them balanced and well-separated.

MERGE integrates three mechanisms: (1) dynamic cluster construction, which creates and updates clusters as new items arrive; (2) real-time occupancy monitoring, which balances cluster sizes and mitigates skew---together forming a single-layer codebook; and (3) fine-to-coarse merging, which builds a hierarchical index for complex business requirements. These components allow MERGE to adapt to evolving item distributions in streaming scenarios.

\subsection{MERGE Framework}

\subsubsection{Dynamic Cluster Construction}

This component incrementally updates clusters from streaming item data, forming a dynamic codebook. We describe the batch-wise update procedure below.

At each step, a batch of items $\mathcal{B}$ is sampled from the item corpus. Each item is represented by a \textit{64-dimensional collaborative embedding} produced by a retriever model serving billions of users. We maintain an initially empty codebook $\mathcal{Q} = \{\bm{q}_k\}_{k=1}^{K}$, where each codeword $\bm{q}_k$ represents a cluster and $K$ evolves over time. Codewords are updated by Exponential Moving Average (EMA)~\cite{EMA}. For each $\bm{q}_k$, we track the EMA sum of assigned item embeddings $\bm{S}_k$ and the EMA count $N_k$.

\textbf{Step 1: Similarity Computation.}
For each item embedding $\bm{e}_i \in \mathcal{B}$, we compute its similarity to all existing codewords and identify the best match:
\begin{equation}\label{eq:sim}
\begin{aligned}
k^*_i &= \mathop{\arg\max}\limits_{{k \in \{1, \dots, K\}}} \text{sim}(\bm{e}_i, \bm{q}_k), \\
s^*_i &= \max_{k \in \{1, \dots, K\}} \text{sim}(\bm{e}_i, \bm{q}_k).
\end{aligned}
\end{equation}
We use cosine similarity, which performs well empirically and yields values in $[-1,1]$ for threshold-based assignment. When embeddings are $\ell_2$-normalized, maximizing cosine similarity is equivalent to minimizing the squared Euclidean distance.

\textbf{Step 2: Matching Decision.} Given a threshold $\tau\in[0,1]$, items are partitioned into matched and unmatched sets:
\begin{equation}\label{eq:decision}
\begin{aligned}
\mathcal{B}^+ &= \{ i \in \mathcal{B} \mid s^*_i \ge \tau\},\\
\mathcal{B}^- &= \{ i \in \mathcal{B} \mid s^*_i < \tau \},
\end{aligned}
\end{equation}
where $\mathcal{B}^+$ and $\mathcal{B}^-$ denote successfully matched and failed items, respectively.

\textbf{Step 3: Cluster Update.}
For each codeword $\bm{q}_k$, we collect its matched items:
\begin{equation}\label{eq:identify}
\mathcal{B}_k^+ = \{ i \in \mathcal{B}^+ \mid k^*_i = k \}.
\end{equation}
Then we update its EMA variables:
\begin{equation}\label{eq:ema}
\begin{aligned}
\bm{S}_k &\gets \gamma \bm{S}_k + (1-\gamma) \sum_{i \in \mathcal{B}_k^+} \bm{e}_i, \\
N_k &\gets \gamma N_k + (1-\gamma) |\mathcal{B}_k^+|,
\end{aligned}
\end{equation}
where $\gamma$ is typically set to 0.99. The codeword is then updated as:
\begin{equation}\label{eq:update}
\bm{q}_k \gets \frac{\bm{S}_k}{N_k}.
\end{equation}
This update enables codewords to adapt to new items while maintaining stability, improving item-to-cluster assignment accuracy.

\textbf{Step 4: Cluster Extension.}
For unmatched items $\mathcal{B}^-$, we construct new clusters using Union-Find~\cite{UF}. This procedure forms connected components under a pairwise similarity threshold without requiring a fixed number of clusters or iterative centroid optimization~\cite{PANDORA,OPA}.

Specifically, we compute pairwise similarities in $\mathcal{B}^-$, connect item pairs above a threshold, and merge connected items into disjoint clusters:
\begin{equation}\label{eq:UF}
\mathcal{U}= \text{UF}(\mathcal{B}^-, \tau^\prime),
\end{equation}
where $\mathcal{U}$ is the set of cluster embeddings obtained by mean-pooling items in each cluster, and $\tau^\prime \in [0,1]$ is the connection threshold.

We retain only clusters whose sizes exceed a threshold $m$:
\begin{equation}\label{eq:rep}
\mathcal{U}^\text{valid} = \{ \bm{u} \in \mathcal{U} \mid \text{size}(\bm{u}) \ge m \},
\end{equation}
where $\text{size}(\bm{u})$ is the number of items in cluster $\bm{u}$.

Let $\mathcal{Q}^\text{zero} \subseteq \mathcal{Q}$ be the reset, zeroed cluster embeddings. To incorporate $\mathcal{U}^{\mathrm{valid}}$, we fill reset entries first and append the rest via the \textit{Fill-Then-Append (FTA)} operator:
\begin{equation}\label{eq:fta}
\mathcal{Q} \gets \mathrm{FTA}(\mathcal{Q}, \mathcal{U}^{\mathrm{valid}}).
\end{equation}
Filling zeroed clusters preserves codebook order and improves stability. The EMA variables for new clusters are initialized accordingly. Since $\mathcal{U}^{\mathrm{valid}}$ comes from unmatched items, the new clusters are naturally distinct from existing codewords, enhancing separation and reducing redundancy.

\textbf{Step 5: Item Recycling.}
Clusters in $\mathcal{U} \setminus \mathcal{U}^\text{valid}$ fail both matching and valid cluster formation. Their items are recycled into the data queue and await future batches for subsequent updates.

\textbf{Tag-Based Batch Formation.}
In streaming settings, item embeddings within the same batch may be highly diverse, causing noisy and unstable cluster updates. To improve homogeneity and accelerate convergence, we use prior tags from multimodal information, covering 100 categories, and construct each batch with items sharing the same tag. This strategy mainly optimizes training when the batch size is not sufficiently large, leading to faster convergence and more stable early-stage cluster formation; it can be removed when the batch size is sufficiently large.

\subsubsection{Real-Time Occupancy Monitoring}

This component runs alongside cluster updates to maintain balanced cluster sizes. Each cluster $\bm{q}_k$ tracks its EMA count $N_k$, and is categorized by two thresholds $\varepsilon_1$ and $\varepsilon_2$:
\begin{equation}\label{eq:status}
\text{status}(\bm{q}_k) \coloneqq
\begin{cases}
\text{underfilled}, & N_k < \varepsilon_1, \\
\text{growing}, & \varepsilon_1 \le N_k < \varepsilon_2, \\
\text{stable}, & N_k \ge \varepsilon_2.
\end{cases}
\end{equation}

Stable clusters remain unchanged. Underfilled clusters are \textit{reset}: their embeddings and EMA variables are zeroed, and associated items are cleared, forming $\mathcal{Q}^\text{zero} \subseteq \mathcal{Q}$. Growing clusters are monitored over $M$ steps and reset if they fail to become stable. Adjusting $\varepsilon_1$ and $\varepsilon_2$ helps balance cluster occupancy, avoiding dominant large clusters and redundant small ones.

\subsubsection{Fine-to-Coarse Merging}

Given the fine-grained codebook $\mathcal{Q}$, we construct a hierarchical codebook by building a coarser codebook $\mathcal{P}$ on top. Initially, $\mathcal{P}=\mathcal{Q}$.

\textbf{Step 1: Merging.}
We define the \emph{affinity score} between $\bm{p}_x$ and $\bm{p}_y$ in $\mathcal{P}$ as:
\begin{equation}\label{eq:aff}
w(\bm{p}_x, \bm{p}_y) = \text{sim}(\bm{p}_x, \bm{p}_y) - \lambda \min(N_x, N_y),
\end{equation}
where $\text{sim}(\cdot, \cdot)$ is cosine similarity, and $N_x,N_y$ are EMA counts. The second term penalizes high-occupancy clusters, encouraging more uniform merging.

At each iteration, we merge the pair with the highest affinity. The merged embedding is computed by count-weighted averaging:
\begin{equation}
\bm{p}_{xy} = \frac{N_x \bm{p}_x + N_y \bm{p}_y}{N_x + N_y},
\end{equation}
with count updated accordingly. The two prototypes are replaced by $\bm{p}_{xy}$, and affinities are recomputed until the target cluster number is reached.

After merging, $\mathcal{P}$ contains merged clusters $\mathcal{P}^\text{mrg}$ and unmerged prototypes $\mathcal{P}^\text{umrg}$.

\textbf{Step 2: Pruning.}
For each merged cluster $\bm{p} \in \mathcal{P}^\text{mrg}$, we assess each prototype $\bm{q}$ using the \emph{silhouette coefficient}~\cite{silhouette}:
\begin{equation}\label{eq:scoeff}
r_{\bm{q}} = \frac{b_{\bm{q}} - a_{\bm{q}}}{\max\{a_{\bm{q}}, b_{\bm{q}}\}},
\end{equation}
where $a_{\bm{q}}$ is the average distance to prototypes in the same merged cluster, and $b_{\bm{q}}$ is the minimum average distance to prototypes in other clusters. The distance is defined as $1-w(\bm{p}_x, \bm{p}_y)$. Prototypes with scores below threshold $r$ are pruned.

\textbf{Step 3: Reconnection.}
The pruned prototypes $\mathcal{P}^\text{prun}$ and unmerged prototypes $\mathcal{P}^\text{umrg}$ are reintroduced into the merging procedure. By iteratively applying merging, pruning, and reconnection, we obtain the desired coarse-grained codebook $\mathcal{P}$.

These steps can be recursively applied for even coarser layers. In this work, we construct only $\mathcal{P}$, which satisfies our online deployment requirements.

\section{Experiment}


\noindent\textbf{RQ1}: How does MERGE compare with baseline indexing methods in accuracy, uniformity, and separation?

\noindent\textbf{RQ2}: Can MERGE's indexing improvements translate into measurable gains in user satisfaction and business metrics?

\noindent\textbf{RQ3}: How do the item--cluster assignments produced by MERGE and baselines differ qualitatively?

\subsection{Offline Evaluation (RQ1)}

We first evaluate MERGE with respect to the three challenges discussed above.

\textbf{Dataset}.
Since MERGE targets large-scale streaming recommendation, we train and evaluate it on platform-specific industrial data rather than public datasets. Following StreamingVQ~\cite{StreamingVQ}, MERGE is trained on a candidate stream where candidates are presented sequentially with equal probability, which is more suitable for indexing than popularity-biased impression streams~\cite{LETTER}. The data contains hundreds of millions of candidates, providing sufficient coverage and diversity.

\textbf{Implementation Details}.
We use a retriever model to provide real-time 64-dimensional item embeddings, incorporating item ID, author features, statistical signals, and content-aware attributes. MERGE is trained with batch size 20,480 to ensure stable cluster assignment. After convergence, the fine-grained codebook contains tens of thousands of valid clusters. Training outputs are written back for subsequent updates. MERGE involves only a few intuitive hyper-parameters, such as assignment/merging thresholds, EMA smoothing, cluster-size regularization, and occupancy-based filtering. \textit{In our experiments, performance is robust to moderate variations of these choices, and detailed sensitivity studies are omitted due to space constraints.}

\textbf{Baseline}.
We compare MERGE with StreamingVQ~\cite{StreamingVQ}, denoted as VQ, which uses a predefined-size single codebook and represents a canonical VQ-based indexing paradigm validated at large scale.

\textbf{Item Sampling}.
We randomly sample items $\mathcal{D}$ from the stream and obtain their assigned cluster embeddings:
\[
\{\, (i, \bm{q}_i) \mid i \in \mathcal{D} \,\}.
\]
Here, $\bm{q}_i$ denotes the corresponding cluster embedding. For MERGE, it is taken from the fine-grained codebook $\mathcal{Q}$. We then evaluate accuracy, uniformity, and separation as follows.

\subsubsection{Accuracy}

\begin{figure}[t]
    \centering
    \includegraphics[width=\linewidth]{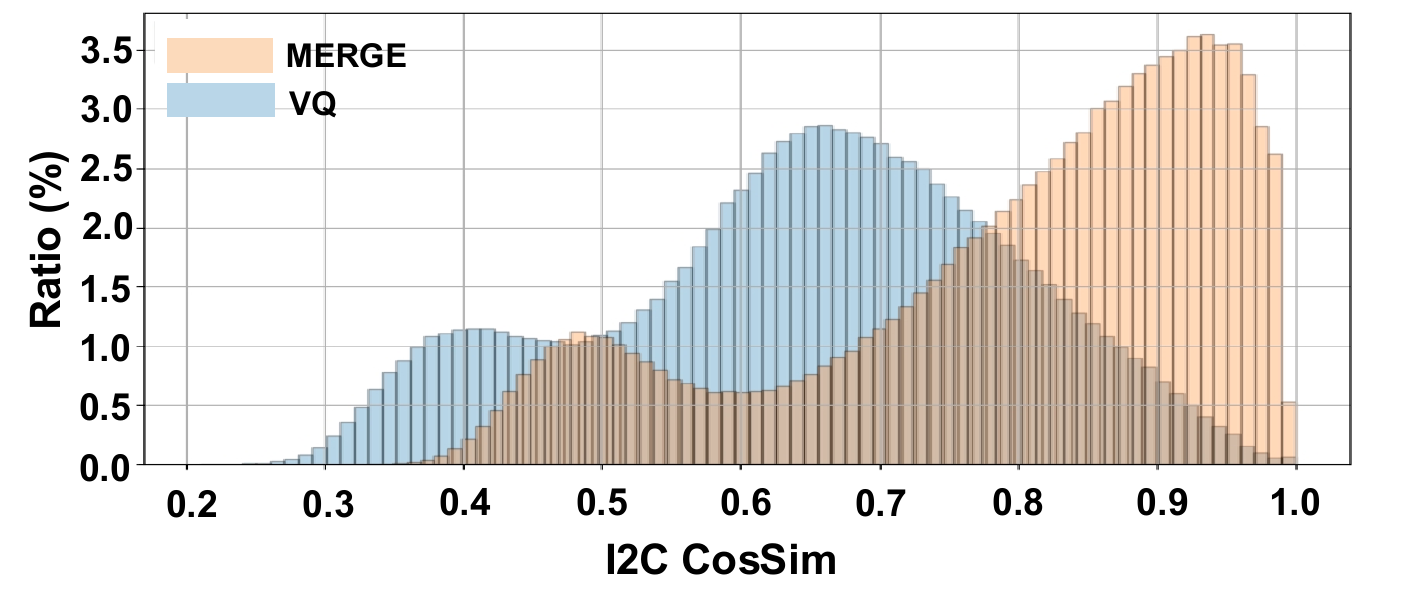}
    \caption{Accuracy comparison between MERGE and VQ, illustrating the item-to-cluster cosine similarity (I2C CosSim) distributions within the codebook.}
    \label{fig:accuracy_comparison}
\end{figure}

We evaluate whether each item is close to its assigned cluster by computing item-to-cluster cosine similarity (\textit{I2C CosSim}). As shown in Figure~\ref{fig:accuracy_comparison}, VQ has a mean similarity around 0.6, while MERGE reaches about 0.9. This indicates that MERGE produces more semantically coherent assignments, mainly because it performs threshold-based matching rather than blind assignment.

\begin{figure}[t]
    \centering
    \begin{minipage}{\linewidth}
        \centering
        \includegraphics[width=\linewidth]{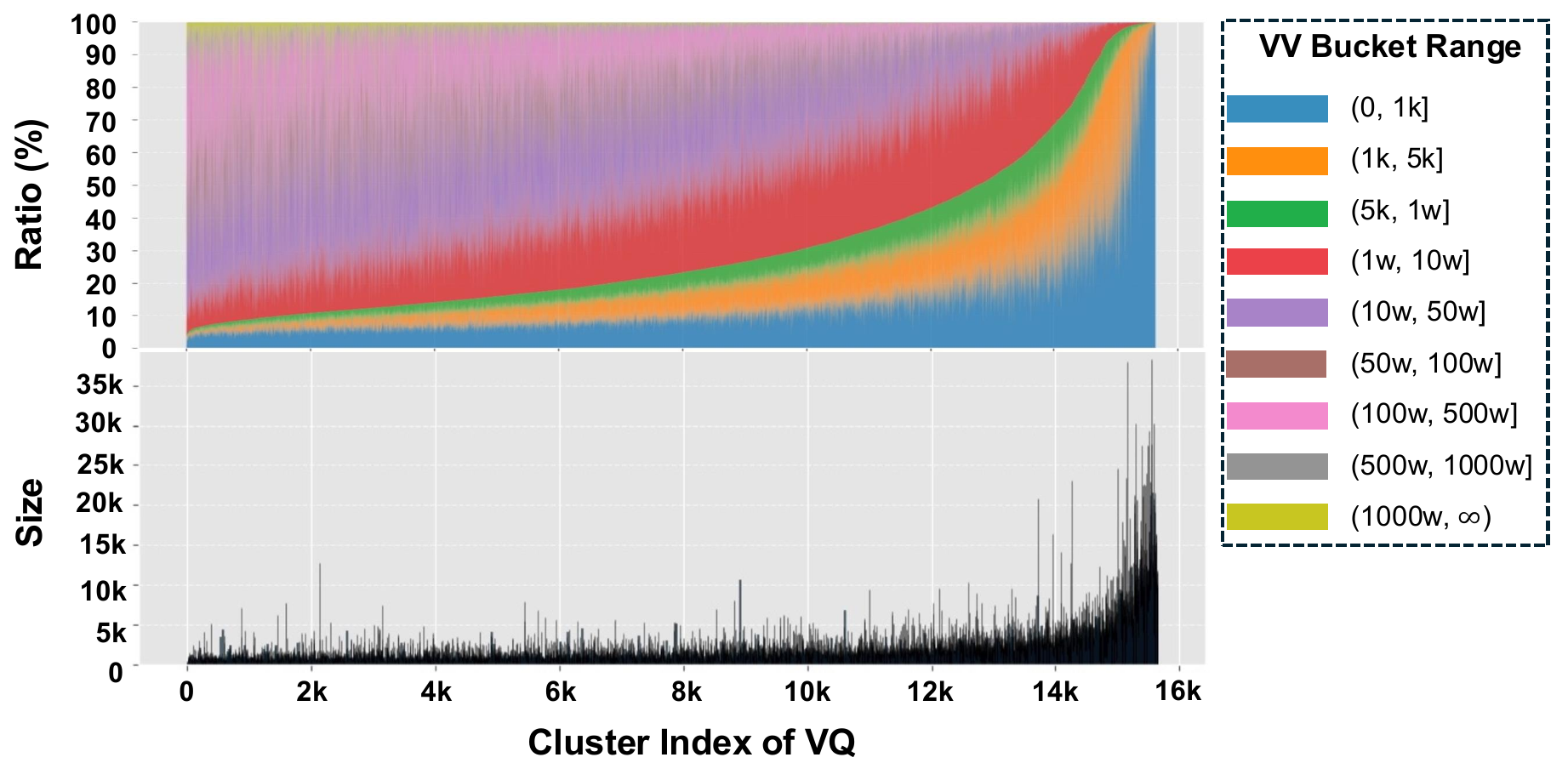}
    \end{minipage}
    \vspace{1em}
    \begin{minipage}{\linewidth}
        \centering
        \includegraphics[width=\linewidth]{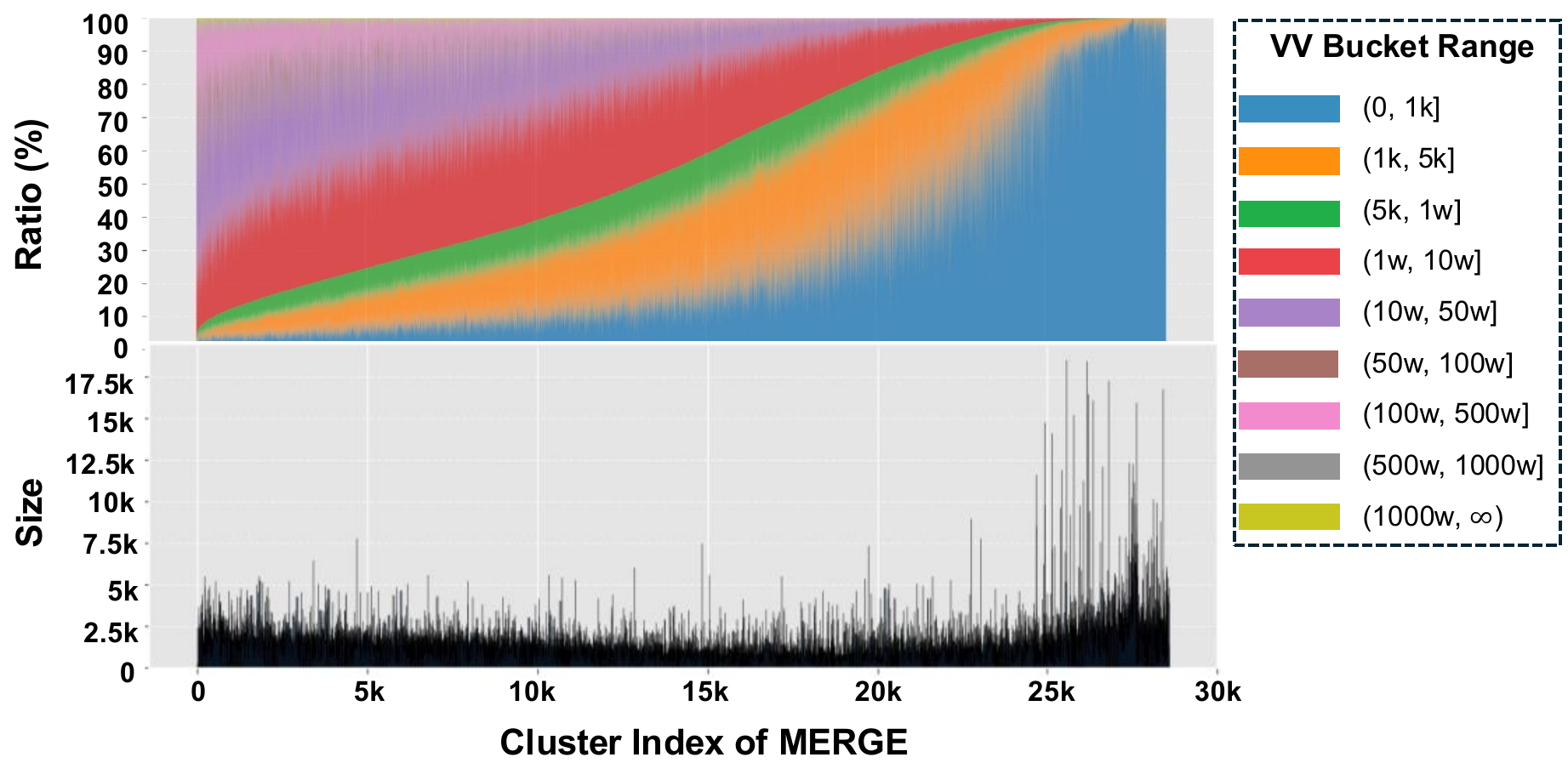}
    \end{minipage}
    \caption{Uniformity comparison between MERGE and VQ, showing cluster sizes and the cumulative distribution of videos within the VV bucket range across clusters.}
    \label{fig:uniformity_comparison}
\end{figure}

\begin{figure}[t]
    \centering
    \includegraphics[width=\linewidth]{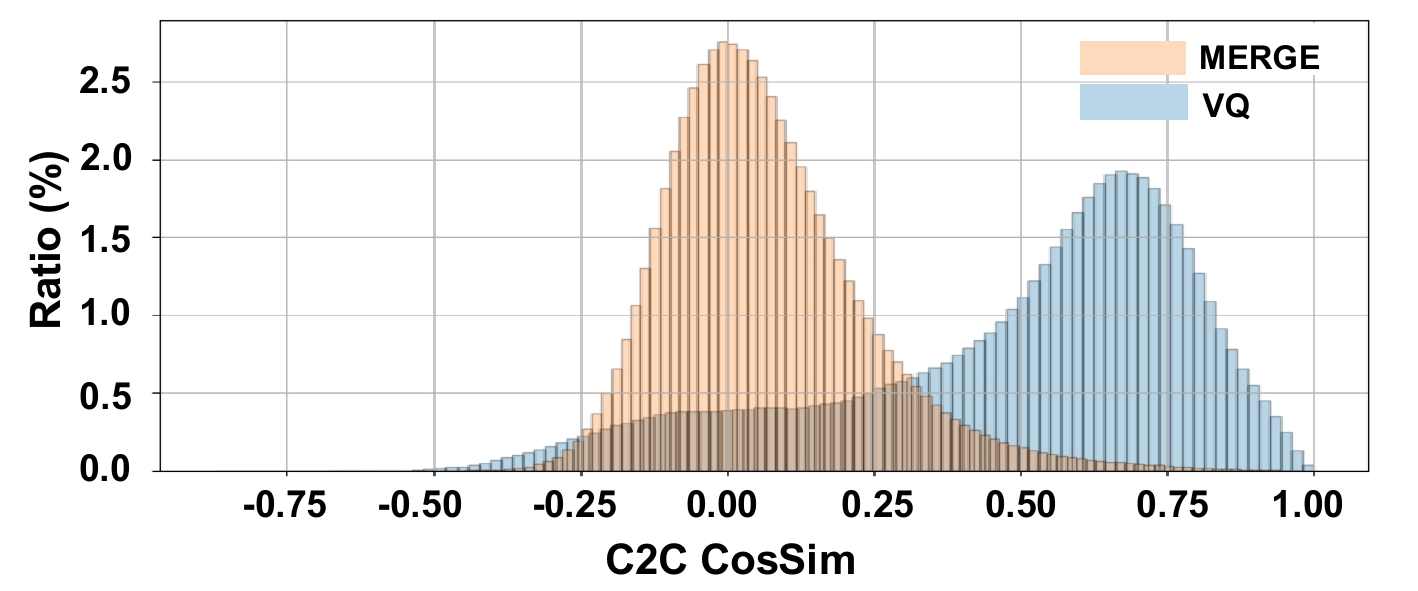}
    \caption{Separation comparison between MERGE and VQ, illustrating the cluster-to-cluster cosine similarity (C2C CosSim) distributions within the codebook.}
    \label{fig:separation_comparison}
\end{figure}

\subsubsection{Uniformity}

We examine cluster sizes, \textit{i.e.}, the number of items assigned to each cluster. For finer analysis, videos are grouped by Video Views (VV), and we visualize how videos in each VV bucket are distributed across clusters. As shown in Figure~\ref{fig:uniformity_comparison}, the largest MERGE cluster contains about 17,500 items, while the largest VQ cluster reaches around 40,000 items, indicating better overall uniformity. Moreover, VQ tends to group low-VV videos together and leaves most clusters dominated by high-popularity items. In contrast, MERGE distributes low-VV videos more evenly, mitigating popularity dominance. This validates the effectiveness of occupancy monitoring and reset operations.

\subsubsection{Separation}

We measure pairwise cluster similarity using cluster-to-cluster cosine similarity (\textit{C2C CosSim}). As shown in Figure~\ref{fig:separation_comparison}, MERGE produces an approximately normal distribution centered near zero, whereas VQ has a much higher mean similarity around 0.6. This demonstrates stronger cluster separation in MERGE, mainly due to its dynamic codebook expansion. Better separation also improves training stability by reducing abrupt item--cluster assignment changes, with theoretical justification provided in Appendix~\ref{sec:stable}.

\subsection{Online Deployment (RQ2)}

After validating MERGE offline, we deploy it in our online recommender system to test whether indexing improvements lead to user satisfaction gains.

\begin{figure}[t]
    \centering
    \includegraphics[width=\linewidth]{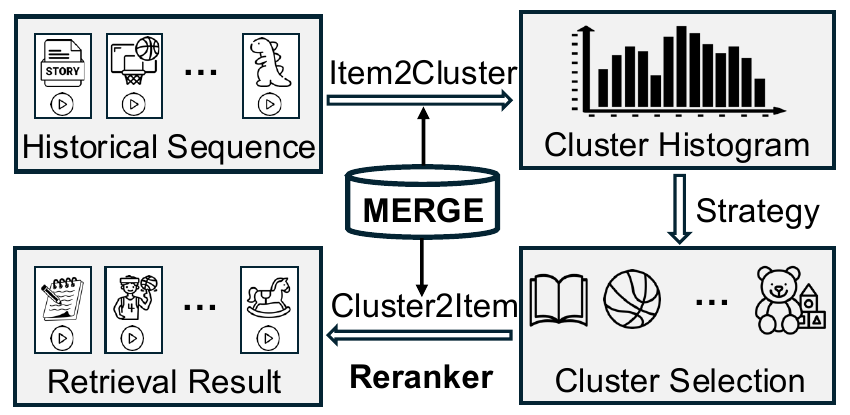}
    \caption{Overview of MERGE deployment during serving, following the Trinity~\cite{Trinity} pipeline. MERGE provides a bidirectional mapping between clusters and items.}
    \label{fig:serving}
\end{figure}

\begin{table}[t]
\caption{Relative changes in core metrics observed during the A/B test. AAD and AAH denote Average Active Days and Average Active Hours, respectively.}
\label{table:global_metric}
\begin{tabular}{ccc}
\hline
\textbf{Aspect} & \textbf{Metric} & \textbf{Relative Change} \\ \hline
\multirow{3}{*}{Engagement} & AAD & +0.0081\% \\
                            & AAH & +0.0546\% \\
                            & WatchTime & +0.1006\% \\ \hline
\multirow{5}{*}{\begin{tabular}[c]{@{}c@{}}VV\\ Stratification\end{tabular}} 
                            & (0,1k] & +1.49\% \\
                            & (1k,5k] & +11.07\% \\
                            & (5k, 10k] & +11.87\% \\
                            & (10k, 100k] & +2.25\% \\
                            & (100k, $\infty$) & -2.65\% \\ \hline
\multirow{4}{*}{\begin{tabular}[c]{@{}c@{}}Freshness\\ Stratification\end{tabular}} 
                            & (0, 2h] & +0.59\% \\
                            & (2h, 12h] & +7.64\% \\
                            & (12h, 24h] & -0.29\% \\
                            & (24h, 72h] & -1.88\% \\ \hline
\multirow{2}{*}{Diversity} & TagNum & +0.22\% \\
                           & TagGini & -0.03\% \\ \hline
\end{tabular}
\end{table}

\begin{table}[t]
\caption{Relative changes in single-path metrics observed during the A/B test.}
\label{table:path_metric}
\begin{tabular}{ccc}
\hline
\textbf{Aspect} & \textbf{Metric} & \textbf{Relative Change} \\ \hline
Survivability & Pass-Through Rate & +45.04\% \\ \hline
\multirow{2}{*}{Contribution} & Output Ratio & +85.99\% \\
                              & Unique Output Ratio & +26.36\% \\ \hline
\multirow{5}{*}{Action} & Staytime & +9.84\% \\
                        & Like & +8.78\% \\
                        & Follow & +17.32\% \\
                        & Share & +7.65\% \\
                        & Comment & +39.43\% \\ \hline
\end{tabular}
\end{table}

\begin{figure}[t]
    \centering
    \includegraphics[width=\linewidth]{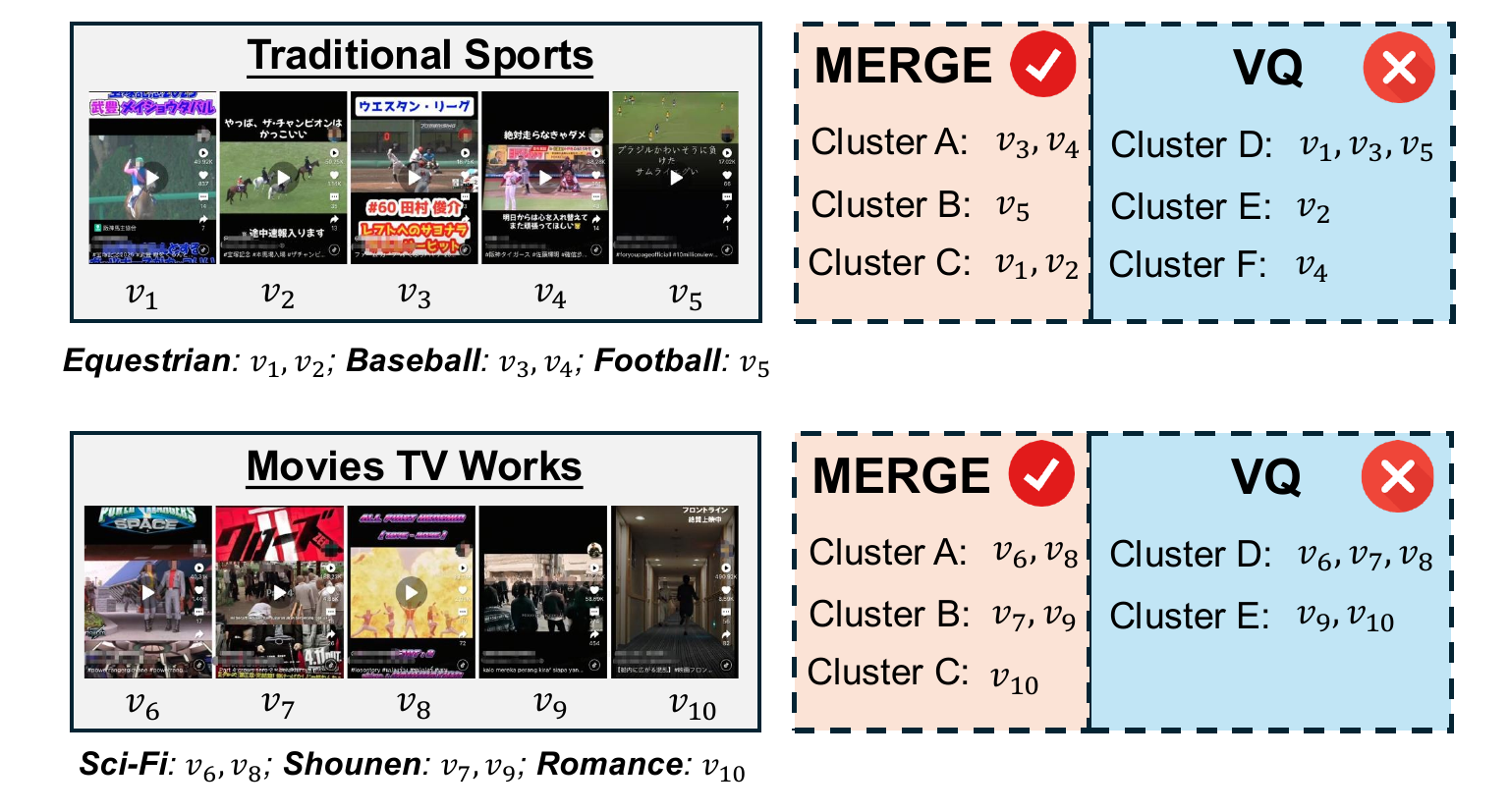}
    \caption{Case study comparing the cluster assignments of retrieved videos under MERGE and VQ.}
    \label{fig:case_study}
\end{figure}

\textbf{Pipeline}.
We follow Trinity~\cite{Trinity}, which uses VQ-based two-level clustering to build statistical interest histograms for modeling multi-interest, long-tail, and long-term user interests. We replace its original VQ indexing with MERGE.
As shown in Figure~\ref{fig:serving}, for each request, we extract the user's long-term behavior sequence and build a histogram over clusters using item--cluster mappings. Clusters are then selected by a predefined strategy, such as top-ranked clusters, following Trinity. Candidate items are retrieved through cluster--item mappings and fed into a reranker, followed by pre-ranking, ranking, and other downstream stages~\cite{DR}.

\textbf{A/B Test}.
We conduct a one-week A/B test with millions of users. The control group uses Trinity with VQ, while the treatment group uses Trinity with MERGE. Raw counts and exact traffic allocation are commercially sensitive, so we report aggregated relative changes.

\subsubsection{Core Metric Analysis}

As an entertainment platform, our primary goal is to improve user engagement. Since DAU is difficult to observe directly under balanced A/B assignment, we follow Trinity and use Average Active Days (AAD), Average Active Hours (AAH), and Watch Time as proxy metrics. We further analyze content composition and diversity. VV stratification measures exposure across popularity levels, freshness stratification measures exposure across publication-time ranges, and diversity is evaluated by TagNum and TagGini.
As shown in Table~\ref{table:global_metric}, MERGE improves all engagement metrics, with statistically significant gains at the 5\% level. Although the relative gains are modest, they imply substantial practical impact at billion-user scale. Stratification results show that MERGE increases exposure of low-VV and recently published content, while also slightly improving diversity.

These gains stem from three factors. First, improved accuracy assigns items to more semantically appropriate clusters, increasing retrieval relevance. Second, better uniformity balances cluster sizes and improves the visibility of low-VV items. Third, stronger separation reduces over-concentration in popular semantic regions, promoting freshness and diversity. Together, these indexing-level improvements propagate through the recommendation pipeline and enhance user engagement.

\subsubsection{Single-Path Metric Analysis}

We evaluate MERGE along a single Trinity retrieval path from three dimensions: survivability, contribution, and action. Survivability measures whether retrieved items survive later ranking stages, corresponding to Pass-Through Rate. Contribution measures the share of final impressions from this path, before and after deduplication. Action measures user feedback, including Staytime, Like, Follow, Share, and Comment.
As shown in Table~\ref{table:path_metric}, MERGE improves all single-path metrics. Retrieved items are more likely to pass through downstream stages, contribute more and more distinct final impressions, and trigger stronger user interactions. These results indicate that single-path retrieval improvements support the overall online gains.

\subsubsection{Efficiency Analysis}

MERGE is significantly more resource-efficient than VQ because it operates on the item corpus rather than massive user interaction logs. In our online setting, VQ requires a distributed cluster of 1,000 instances with 4 CPU cores each, whereas MERGE achieves superior indexing quality on a single 30-core instance. This greatly reduces synchronization overhead and infrastructure costs.

\subsection{Case Study (RQ3)}

We present a real-world case study of item--cluster mappings. For each request, we retrieve about 1,000 candidate videos and use tag-based filtering to retain around 100 videos of the same theme. We then rank the assigned clusters by frequency and inspect their videos. The selected themes include \textit{``Traditional Sports''} and \textit{``Movies TV Works''}. For a fair comparison, both methods use codebooks of approximately 20,000 clusters. As shown in Figure~\ref{fig:case_study}, MERGE produces fewer and more semantically coherent clusters, whereas VQ yields fragmented and noisy assignments. For \emph{Traditional Sports}, MERGE groups baseball videos ($v_3, v_4$) and equestrian videos ($v_1, v_2$) into their respective clusters. For \emph{Movies TV Works}, it similarly groups Sci-Fi videos ($v_6, v_8$) and Shounen videos ($v_7, v_9$). In contrast, VQ scatters related videos across clusters and introduces cross-category contamination.
These cases demonstrate MERGE's stronger discriminability and assignment accuracy, stemming from its paradigm shift from matching clusters to generating clusters. Accuracy aligns items with proper clusters, while uniformity and separation preserve fine-grained partitioning. Even when enlarging the VQ codebook, these issues remain, suggesting that the limitation lies in its underlying mechanism rather than scale.
\section{Related Work}

Efficient item indexing is essential for large-scale recommendation, as exhaustive corpus scanning is infeasible. Classical ANN methods such as Product Quantization (PQ~\cite{PQ}), Navigable Small World (NSW~\cite{NSW}), and Hierarchical NSW (HNSW~\cite{HNSW}) enable fast approximate retrieval through quantization or graph-based search. Tree-based methods, including TDM~\cite{TDM} and JTM~\cite{JTM}, organize items hierarchically for coarse-to-fine candidate search, while Deep Retrieval~\cite{DR} avoids the Euclidean-space assumption of typical ANN methods and performs efficient path-based indexing. Recently, StreamingVQ~\cite{StreamingVQ} applies VQ to streaming item indexing and achieves notable online gains. StreamingVQ is the closest baseline because it consumes streaming item representations and uses the same one-codebook interface. In contrast, MERGE establishes a cluster-generation paradigm for indexing, with both theoretical and empirical evidence showing its advantages over traditional VQ-based methods.

In generative recommendation~\cite{GenRec_survey}, item indexing is often studied as item tokenization. TIGER~\cite{TIGER} uses RQ-VAE to quantize item text embeddings into token sequences, reformulating recommendation as sequence generation. OneRec~\cite{OneRecv2} adopts a similar strategy and deploys RQ-KMeans at scale in Kuaishou. Most existing methods~\cite{BLOGER,ETEGRec,LETTER,GRID,COBRA,MMQv2,OneSearch,CAT-ID} still rely on VQ, whose inherent paradigm limitations make them less suitable for large-scale streaming recommendation.

\section{Conclusion}

In this work, we investigated the limitations of existing item indexing methods and identified three challenges in industrial streaming recommendation: poor assignment accuracy, imbalanced cluster occupancy, and insufficient cluster separation. We proposed MERGE, an item indexing framework that constructs clusters from scratch, dynamically monitors occupancy, and builds a hierarchy via fine-to-coarse merging. MERGE provides a promising indexing paradigm for large-scale streaming recommendation.

MERGE still has limitations. Low-VV videos remain challenging, as shown in Figure~\ref{fig:uniformity_comparison}. The framework is also more complex than VQ-based methods, and its current deployment is limited to traditional retrieval. Future work will explore multimodal item embeddings, more efficient hierarchical construction, and the extension of MERGE to generative retrieval and reasoning~\cite{IntuRec,LatentR3}.

\appendix

\section{Theoretical Stability Analysis}
\label{sec:stable}

Let $\mathcal{Q}=\{\bm{q}_k\}_{k=1}^K$ be a codebook with normalized codewords,
\textit{i.e.}, $\|\bm{q}_k\|=1$. Suppose item $\bm{e}_i$ is assigned to
$\bm{q}_k$, such that
\begin{equation}
\bm{q}_k^\top\bm{e}_i
\ge
\bm{q}_j^\top\bm{e}_i,
\qquad \forall j\neq k.
\end{equation}

Under a perturbation $\bm{\delta}$ with $\|\bm{\delta}\|\le\epsilon$, the
assignment is preserved if
\begin{equation}
(\bm{q}_k-\bm{q}_j)^\top\bm{e}_i
\ge
(\bm{q}_j-\bm{q}_k)^\top\bm{\delta},
\qquad \forall j\neq k.
\end{equation}
By Cauchy--Schwarz, a sufficient condition is
\begin{equation}\label{eq:stable_condition}
(\bm{q}_k-\bm{q}_j)^\top\bm{e}_i
\ge
\epsilon\sqrt{2(1-\bm{q}_k^\top\bm{q}_j)},
\qquad \forall j\neq k,
\end{equation}
where we use
$\|\bm{q}_k-\bm{q}_j\|^2=2(1-\bm{q}_k^\top\bm{q}_j)$.

To relate stability to codeword separation, decompose
\begin{equation}
\bm{e}_i=\alpha\bm{q}_k+\bm{\Delta},
\qquad
\alpha=\bm{q}_k^\top\bm{e}_i>0,
\qquad
\bm{q}_k^\top\bm{\Delta}=0.
\end{equation}
Then,
\begin{equation}
(\bm{q}_k-\bm{q}_j)^\top\bm{e}_i
=
\alpha(1-\bm{q}_k^\top\bm{q}_j)
-\bm{q}_j^\top\bm{\Delta}
\ge
\alpha(1-\bm{q}_k^\top\bm{q}_j)-\|\bm{\Delta}\|.
\end{equation}
Consequently, Equation~\eqref{eq:stable_condition} holds if
\begin{equation}
\alpha(1-\bm{q}_k^\top\bm{q}_j)-\|\bm{\Delta}\|
\ge
\epsilon\sqrt{2(1-\bm{q}_k^\top\bm{q}_j)},
\end{equation}
which is equivalent to
\begin{equation}\label{eq:stable_final}
\bm{q}_k^\top\bm{q}_j
\le
1-
\frac{
  \left(\epsilon+
  \sqrt{\epsilon^2+2\alpha\|\bm{\Delta}\|}\right)^2
}{2\alpha^2}.
\end{equation}

Therefore, greater codeword separation provides a larger stability margin
against bounded perturbations, supporting the empirical observation in
Figure~\ref{fig:separation_comparison}.

\section*{GenAI Usage Disclosure}

The authors used generative AI tools for language polishing, grammar correction, and improving the clarity and conciseness of the manuscript. All technical content, experiments, analyses, and conclusions were developed and verified by the authors, who have reviewed the resulting text and take full responsibility for the final manuscript.

\bibliographystyle{ACM-Reference-Format}
\balance
\bibliography{8_reference}


\end{document}